\documentclass[pre,letterpaper,twocolumn,final,superscriptaddress,floatfix]{revtex4}

\usepackage{graphicx,psfrag,amsmath,amssymb,amsfonts,bbm,latexsym,color,dcolumn,bm,mathbbol}
\usepackage[framemethod=tikz]{mdframed}
\usepackage{hyperref}
\usepackage{xfrac} 
\usepackage{ifthen}

\usepackage{amsmath,amssymb}             
\usepackage[french, english]{babel}

\newcommand{\refn}[1]{Eq. (\ref{#1})}
\newcommand{\secn}[1]{Sec. \ref{#1}}
\newcommand{\ov}[1]{\overline{#1}}

\newcommand{\Zi}{Z_i}

\newcommand{\Ii}{I_i}

\newcommand{\volume}{S}
\newcommand{\nbpixel}{F}

\newcommand{\estpixel}{\widehat{F}^{(t)}}
\newcommand{\estpixeltheta}{\widehat{F}^{(t_{\theta})}}

\newcommand{\estvolu}{\widehat{\volume}^{(t)}}

\begin{document}

\title{Trade-offs between spatial and temporal resolutions in stochastic super-resolution microscopy techniques}

\author{Jean-Fran\c cois Rupprecht}
\affiliation{Mechanobiology Institute, National University of Singapore, 5A Engineering Drive 1, 117411 (Singapore).}
\email{mbijr@nus.edu.sg}

\author{Ariadna Martinez-Marrades}
\affiliation{Sorbonne Universit\'es, UPMC Univ Paris 05, Paris (France).}

\author{Rishita Changede}
\affiliation{Mechanobiology Institute, National University of Singapore, 5A Engineering Drive 1, 117411 (Singapore).}
\email{mbijr@nus.edu.sg}

\author{Gilles Tessier}
\affiliation{Sorbonne Universit\'es, UPMC Univ Paris 05, Paris (France).}

\begin{abstract}
Widefield stochastic microscopy techniques, such as PALM or STORM, rely on the progressive accumulation of a large number of frames, each containing a scarce number of super-resolved point images. We justify that the redundancy in the localization of detected events imposes a specific limit on the temporal resolution. Based on a theoretical model, we derive analytical predictions for the minimal time required to obtain a reliable image at a given spatial resolution, called image completion time. In contrast to standard assumptions, we find that the image completion time scales logarithmically with the image size to spatial resolution volume ratio, which is the hallmark of a random coverage problem. We discuss how an increased background noise can affect the image completion time and impact the latter scaling. We finally propose a method to estimate the risk that the image reconstruction is not complete. Our results provide a theoretical framework to quantify in real-time the pattern detection efficiency with applications to structural imaging in $1$, $2$ or $3$ dimension.
\end{abstract}

\date{\today}

\maketitle

\section{INTRODUCTION}

Optical microscopy is a convenient tool to study biological processes, but its resolution is fundamentally limited by Abbe's diffraction. The image of a point source is a pattern whose size is comparable to the optical wave-length ($\sim 250 \, \mathrm{nm}$), hence source points separated by a distance smaller than a wavelength are hardly distinguishable \cite{McCutchen1967}. Electron microscopy provides a higher spatial resolution ($\sim 1 \, \mathrm{nm}$) but at the cost of a more complex sample preparation which is incompatible with \textit{in vivo}-imaging \cite{Koster2003}. The recently developed super-resolution imaging techniques aim at combining the best of these two worlds. Using these techniques, spatial resolution as low as $10 \, \mathrm{nm}$ have been achieved for imaging biological cell structures. However, their applicability to the study of dynamical biological processes is limited by their long acquisition times \cite{betzig_imaging_2006, Liu2015}.

Though relying on different optical probes, the super-resolution techniques known as PALM (Photoactivation Localization Microscopy) or STORM (Stochastic Optical Reconstruction Microscopy) rely on a common principle: sources that lie within the same diffraction-limited volume are separated by a sequential activation process, which introduces a temporal separation between source points \cite{Schermelleh2010}. Within each frame, a small and random fraction of probes is activated by illumination. This sparse subset of randomly activated probes is imaged to produce a frame. Then, finding the centroid of each diffraction patterns leads to a set of coordinates, having a nanometer-level precision \cite{McCutchen1967,Betzig1995}. Merging all the single-molecule positions obtained on successive frames produces the final image.  Since only a small fraction of probes is imaged per frame, a certain number of frames is required in order to obtain a reliable reconstructed image. Multiplying this number by the typical acquisition time of frames (typically in the $10-100 \mathrm{ms}$ range), we obtain the minimal time, denoted $T$, to obtain an image at a nanometer-scale resolution. A typical reported value is $T \sim 30 \, \mathrm{min}$ for a whole cell imaging at a $10 \, \mathrm{nm}$ resolution \cite{Liu2015}. This value is too large to study many dynamical processes that occur in living cells, such as the contraction of acto-myosin units \cite{Wolfenson2015}, reorganization of focal adhesion complexes \cite{Bertocchi2013} or protein cluster formation within the plasma membrane \cite{Changede2015a,Biswas2015}. 

In addition, stochastic microscopy is prone to localization errors. These errors may either originate from overlapping spread functions or from emission outside of the region of interest. However, given a set of localized observations, one can generally assume that spurious detections corresponds to regions with low count-density, e.g. the density-based spatial clustering of applications with noise (DBSCAN) algorithm operates noise filtering by eliminating observations whose nearest neighbors are further than a prescribed threshold distance \cite{ester1996density}. Here, we consider the reconstruction criteria in which a region of space is assumed to belong to the region of interest if and only if it has collected at least a number $r$ of observations during the duration of the experiments $T$. 
If the number of observations is insufficient, the reconstructed structural image displays voids within the region of interest (ROI) -- parts of the ROI are assigned to the background noise. We refer to these voids as \textit{stochastic aberration}. This leads to the following two formulations of the main question of the present paper: What should be the minimal acquisition time in order to reliably discriminate between the region of interest from the rest of the field of view? How can we reliably discriminate whether a hole in the reconstructed super-resolved image is a genuine gap in the structure rather than an aberration due to a lack of observations? 


It is generally thought that in widefield stochastic techniques, such as PALM, the imaging time $T$ is solely controlled by the density (denoted $\rho$) of activated fluorophores per frame and by the spatial resolution ($\sigma$) following the relation $T \sim 1/(\rho \sigma)$. The latter relation does not depend on the total size of the field of view ($S$), which represents a subtantial advantage of stochastic techniques over deterministic one.  In raster-scan-based techniques, e.g. STimulated Emission Depletion (STED), the resolution is exempt of stochastic aberration but the acquisition time $t$ increases linearly with the size of the field of view \cite{willig_nanoscale_2006}. 

In this paper, we argue that the imaging completion time $T$ should be expected to depend on the size of the field of view,  due to the \textit{random} localization process that results in an uneven spatial distribution of events. Based on a stochastic model, we derive the relation:
\begin{align} \label{eq:main_message}
T \underset{\sigma/\volume \ll 1}{\sim} \frac{1}{\sigma \rho} \left\lbrace \ln\left(\frac{\volume}{\sigma \theta}\right) + ( r + \gamma_D) \ln\left[\ln\left(\frac{\volume}{\sigma} \right) \right] \right\rbrace,
\end{align}
which means that the trade-off between the spatial ($\sigma$) and temporal ($T$)  resolutions depends (i) on the ratio of the size of the field of view to the desired spatial resolution (ii) on the necessity to separate the ROI from the noisy background, via a
minimal number of redundant observations $r$ that is an increasing function of the background noise intensity,
(iii) on the risk of an incomplete coverage of the ROI (i.e. of stochastic aberrations) via the $5\%$--centile parameter $\theta = 0.95$ and finally (iv) on the dimensionality $D \in \left\lbrace 1,2,3 \right\rbrace$ of the ROI via the constant $\gamma_D$. The prefactor $\ln(\volume/\sigma)$ in \refn{eq:main_message} can be significantly larger than $1$, eg. a cell of extension $\volume = 10^{3}\, \mu\mathrm{m}^2$ contains $10^{7}$ squares of area $\sigma = 10^{-4} \, \mu\mathrm{m}^2$ (ie. a typical size for an Abel diffraction pattern), which leads to $\ln(\volume/\sigma) = 16$. 

The result \refn{eq:main_message} applies to experimental situations in which a high reconstruction fidelity is needed.  Obtaining a complete image reconstruction can be of critical interest in structural reconstructions, e.g. when evaluating the integrity of a DNA segment \cite{Alexeyev2013} or the tensegrity of the actin network within a cell \cite{Zhang2016}. Indeed, a broken actin filament cannot support tension, similarly to a nano-wire which cannot conduct current when it is cut in two. Mind that the logarithmic scaling with the image size stems also holds for a \textit{near} complete image reconstruction (see Sec. \ref{sec:near_complete}). However, in some other experimental contexts in which a high fraction of missed pixels in the reconstructed image is tolerable, we show that the image time should be expected to scale linearly with size of the sample (see Sec. \ref{sec:coupon_not_hold}). 

We tested the applicability of \refn{eq:main_message} on experimental localization sequences. We find that \refn{eq:main_message} is no longer valid at high level of background noise. This is expected since \refn{eq:main_message} does not hold when the minimal number of redundant observations $r$ is larger than a critical value $r_c = \ln(S/\sigma)$. In the regime $r \gg r_c$, we find that the image time behaves as $T \sim r/(\rho \sigma)$. 

\begin{figure}[t!]
\includegraphics[height=8cm]{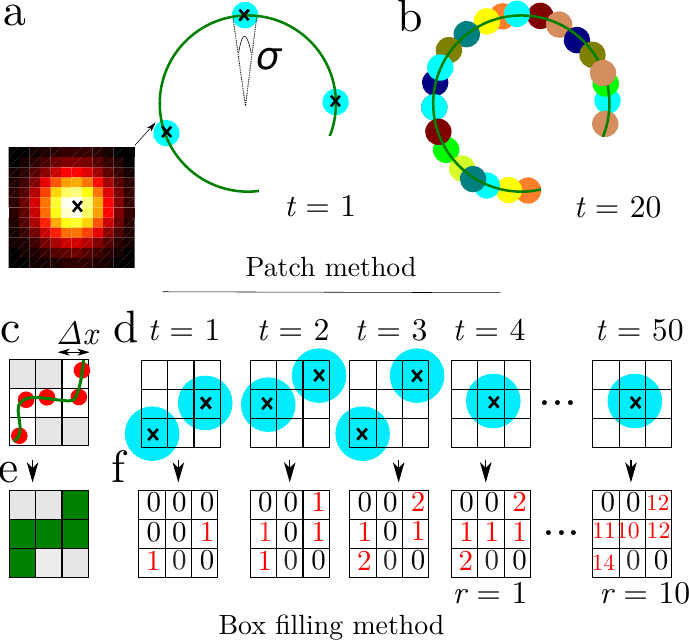}
  \caption{(Color online) Structural reconstruction by stochastic super-localization microscopy. Probes (colored dots) are bound to a structure of interest (green line). (a--b) Circular patches representation: (a) Upper left inset: Abel diffraction pattern observed in a CCD camera. The super-resolution algorithm yields a set of coordinate corresponding to the center of the pattern (black cross). In the patch method representation, each point coordinate is represented by a disk with a radius $\sigma$ that is proportional to the uncertainty of the super-localization procedure (blue disk). (b) Patches accumulate with the acquisition time, eventually covering the whole structure of interest (patches are represented by different colors for separate time frames).
(c--f) Box-filling representation, which leads to a density map in terms of a number of accumulated events per pixel. (c) The field of view is divided into $N = 9$ pixels among which $\nbpixel = 5$ pixels contain probes. (d) A sequence of frames (blue circle: size of the Abel pattern). (e) Target image. (f) Map of the cumulative number of observations $M^{(t)}_j$, for each pixel $j$ and for each frame $t$. Complete image completion (with $r\geq 1$) is obtained after $t=4$ frames. At $t=50$, all pixels have been observed at least $r=10$ times.
  }
\label{Fig1}
\end{figure}

The paper is organized as follows. We first present the two experimental setups: (i) a PALM setup
and (ii) a Total Internal Reflection Microscopy  (TIRM) experiment in which we measure the scattered light from Brownian nano-particles at the surface of a two-dimensional sample. We then define two image rendering schemes, called patch and box-filling methods (see \ref{Fig1}). We then prove the relation \refn{eq:main_message} and we show its connexion to the coupon-collector problem \cite{Paul1961,Feller1968,Stanley1990}. Therefore, we refer to the result of  \refn{eq:main_message} as the \textit{coupon-collector scaling}. 

We then consider the robustness of the coupon-collector scaling for several image completion requirements, and in particular the effect of correlations between successive frames. This case is particularly motivated by the Brownian setup, in which the escapes and returns of the Brownian particles within the detection zone leads to temporally correlated scattering events between successive frames. We point out that there is a close analogy between the gold nanoparticle experiments and PALM techniques relying on organic dyes whose blinking statistics exhibit time-correlations  \cite{Annibale2011}. We recall that bleaching refers to an irreversible transition of a probe to an inactive state \cite{Schermelleh2010}. The analogy holds both on correlated blinking events -- which corresponds to the correlated returns of the Brownian particles to the illuminated region -- and on bleaching events -- which corresponds to the escape of the Brownian particle far from the illuminated region.

We conclude our article by presenting a procedure by which, in real-time during the acquisition, we can estimate the risk that the image is prone to stochastic aberrations.

\section{METHODS} \label{sec:methods}

\begin{figure}[t!]
\includegraphics[height=8cm]{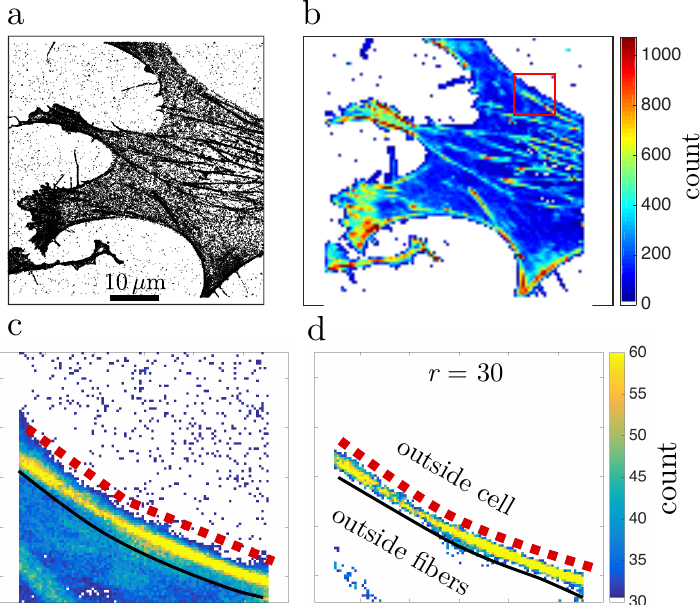}
  \caption{(Color online) PALM imaging of the actin mesh within a fibroblast cell. (a) Image obtained by representing detection events as circles of radius $5\mathrm{nm}$ (i.e. patch method).  (b) Density plot, in which the field-of-view is divided into $100 \times 100$ pixels of width $x \ \mathrm{nm}$.  (i.e. box-filling method).  (c-d) Zoom on the cell boundary plot, which reveals three regions defined according to the density of observations: low (above red line), intermediate (below black line) and high (between the two lines). (d) Low density filtering: only pixels which have collected more than $r = 30$ observations are represented, to isolate the region of interest (i.e. actin fibers at the boundary). Mind that $r$ represents the number of redundant observation required to separate a region of interest from a noisy background.
  }
\label{Fig2}
\end{figure}

\subsection{PALM experiments} \label{sec:method:PALM_experiments}

We analyzed the sequence of localization events from two sets of samples: (i) silane sample with quasi-uniform sampling in fluorophores, and (ii)  fibroblast cell with tagged actin structure.
In particular, we illustrate our noise removal procedure in Fig. \ref{Fig2}, by requiring a minimal number of $r = 30$ events. Details on the PALM experiments are provided in the SI \cite{SI}.

\subsection{TIRM experiments}  \label{sec:method:Brownian_experiments}
In a recent work \cite{Martinez-marrades2014}, we presented a new stochastic imaging technique to map an electromagnetic field with a nano-scale resolution using light-scattering Brownian particles as local probes of the field intensity. The Brownian motion of the scatterers eventually lead to a full coverage of the imaged field. Following \cite{Martinez-marrades2014}, we consider the imaging problem of an evanescent wave created by a Total Internal Reflection Microscopy setup. In this setup, we consider that the optical intensity of the electromagnetic field can be modelled as
\begin{align} \label{eq:io}
I(x, y, z) = I_0(x, y) \exp(- z/\beta(x, y)),
\end{align}
where $\beta$ is the penetration length of the field, and $I_0$ is proportional to the optical intensity of the field at the surface -- with a proportionality constant related to the scattering cross section of the particles. 

In principle, in most situations of interest, both quantities $\beta$  and $I_0$ can vary with the location $(x,y)$ on the surface. In this context, the term image acquisition refers to the determination of the field intensity $I_0$ and $\beta$. However, as a first test of the method, the experimental data set from \cite{Martinez-marrades2014} corresponds to a situation where both $I_0$ and $\beta$ are homogeneous within the whole field of view. We detail a procedure that leads to the determination of $I_0$ and $\beta$ in the SI \cite{SI}.

\subsection{Two image rendering methods} \label{sec:method:rendering}

Super-resolution techniques rely on the localization of the center of diffraction spots, which provides a set of points. However, a spatial extention needs to be attributed to each point to obtain an image that is readable to the human eye. In the following, we will be interested in the two following image rendering methods: (i) the box-filling method (BFM), which is adapted to a density image representation \cite{Cox2011a} and (ii) the patch method (PM), which is associated to a pointillist representation \cite{Triller2005}. 

The BFM considers the structure of interest as tessellated into $\nbpixel$ square pixels of equal area, which can therefore be expressed as the ratio of the total volume by the resolution volume: $\nbpixel = \volume/\sigma$. Each new event falls within a specific pixel, thereby increasing by one the cumulative number of observations of this pixel. This method is naturally adapted to measure the densities. Though we employ the term pixel in the following, our method also applies to 3D imaging problems in which $\nbpixel$ refers to the number of voxels within the structure of interest \cite{Juette2008,Shtengel2008,Hajj2014,Galland2015}.

In the other hand, the PM associates to each event a surrounding extension, characterized by the quantity $\sigma$, which is either a length (1D), an area (2D) or a volume (3D). Generally, the spatial extension is chosen to correspond to the spatial uncertainty associated to the localization procedure (e.g. a few nanometers, \cite{Triller2005}). The image completion time is related to the minimal number of patches required to cover the structure of interest.  


\subsection{Statistics of events}

We assume that fluorescent events are distributed according to a homogeneous Poisson process, such that the probability density $dP$ that an event occurs in an infinitesimal space of volume $d s$ reads  $dP = \rho d s$ \cite{Hall1988}. We now consider a regular domain of volume $\volume$ within a $D$ dimensional space, in which we assume a constant density of fluorophore $d$. Furthermore, we assume that at each frame, only a fraction $f$ of fluorophores are detected. The number of detected fluorescence events after one frame, denoted $N^{(1)}$, is a Poisson process of density $\rho = f d$; hence $P\left[ N^{(1)} = n \right] = \exp(- \rho \volume) (\rho \volume)^{n}/n!$. If $\volume = A$ refers to the volume of the Abel diffraction pattern, the mean number of fluorescence events per frame  $\rho A$ should be lower than $1$ in order to limit the risk of overlapping point spread functions. Typically $A = (10^{2} \, \mathrm{nm})^{D}$ \cite{Schermelleh2010}, hence $\rho < 10^{-2D} \mathrm{nm}^{-D}$. In the case of membrane ($D=2$) with fluorophore density $d = 10^{4} \, \mu \mathrm{m}^{-2}$, the corresponding maximal fraction of activated fluorophores should be $f < 10^{-3}$. After a number $T$ of frames, the total number of collected events is distributed according to a Poisson distribution, with $P\left[ N^{(T)} = n \right] = \exp(- \rho \volume T) (\rho \volume T)^{n}/n!$.  


We finally assume that the density of events $\rho$ is time-independent, hence neglecting the progressive bleaching of fluorophores \cite{Annibale2011}. Our time-independent assumption corresponds to two situations, in which either (i) the total number fluorophores per elementary resolution volume remains large compared to the number bleached fluorophores, or (ii) if the activation laser is increased as a function of time in order to balance the effect of bleaching.


\subsection{Estimation of the structure size}

In many situations, the volume of the structure is either completely unknown \textit{a priori}, or can only be partially inferred - e.g. by assuming randomly oriented linear order. Within the BFM, we show (see SI, Sec. \ref{sec:app:Festimator} \cite{SI}) that the maximum likelihood estimator of the number of relevant pixels $\nbpixel$ corresponds to the quantity:
\begin{align} \label{eq:Festimator}
\estpixel = \sum^{\nbpixel}_{j = 1} \min \left( M^{(t)}_j, 1 \right),
\end{align} 
where $M^{(t)}_j$ is the cumulative number of measures of the pixel $j$, e.g. $M^{(t)}_j = 0$ if the pixel $j$ has never collected any event up to time $t$ and $M^{(t)}_j \geq 1$ if the pixel has been observed at least once up to time $t$ (see Fig. \ref{Fig1}). Similarly, within the patch-method framework, the maximum likelihood estimator of the structure volume consists in the covered volume at the time $t$. These two estimators are biased, as they tend to underestimate the structure volume.


\subsection{Mathematical definition of the image completion time}
We call \textit{image completion time} the minimal number of frames required to obtain a complete image of the region of interest. The term complete refers to the condition that every pixel or point (among those that should be observed) has been covered at least a certain number of times, denoted $r \geq 1$. More precisely,  the image completion time $T$ is the random variable (called stopping time) that corresponds to the minimal time $t$ such that $\min_{j}\left(M^{(t)}_j\right) = r$; where $j \in \left[1, \ldots \nbpixel \right]$ in the BFM framework, or $j$ refers to any point within the volume of interest in the PM framework. We will be mainly interested in the centile of $T$, denoted $t_\theta$ and defined as:
\begin{align} \label{def:t_theta}
\mathbb{P}\left[ T \leq t_\theta \right] = \mathbb{P}\left[\min_{j}\left(M^{(t_\theta)}_j\right) \geq r \right] = 1-\theta,
\end{align}
where $\theta$ is the tolerated risk. To summarize, the quantity $t_{0.05}$ refers to the minimal number of frames that guarantees, with $95\%$ probability, that there is no stochastic aberration within the reconstructed ROI image. 

\subsection{Simulations}

Both in the BFM and PM frameworks, the volume of the region of interest is tessellated into a grid of elementary squares. In the BFM, each event covers a single elementary square; while in the PM, each patch $\sigma$ covers a square matrix of elementary squares. In both frameworks, we generate a large sample of coverage events and we analyse the resulting distribution of coverage times using Matlab's \textit{prctile} function.

\section{RESULTS AND DISCUSSION}


\subsection{The image completion time follows a coupon-collector scaling } \label{sec:imagingtimer1}

We first derive the main result of \refn{eq:main_message} in the case of the BFM with no time-correlation between frames. Here, we assume that the value of the total number of pixels $\nbpixel$ is known.  Under the assumption that detection events occurring in separate pixels are independent, the probability that exactly $M$ pixels have been observed at least once ($r=1$) reads:
\begin{align}  \label{eq:subset}
\mathbb{P}\left[ \estpixel  = M \right] = {\nbpixel \choose M}  p_0^{(M-\nbpixel)t} \left(1-p_0^{t}\right)^{M},
\end{align} 
where $p_0 = 1- p_1$, and $p_1 = \rho \sigma$ is the probability that an event occurs in a given pixel and at a given frame. In particular, the probability that the estimator $\estpixel$ is equal to its target value $\nbpixel$ reads:
\begin{align} \label{eq:Ftheta}
\mathbb{P} \left[ \estpixel = \nbpixel \right] = \left(1-(1- p_1)^{t}\right)^{\nbpixel}.
\end{align} 

We now determine the centile of the image completion time, defined according to \refn{def:t_theta} as the solution of the equation $\mathbb{P}\left[ \estpixeltheta = \nbpixel \right]  = 1 - \theta$, hence 
$t_\theta = \ln\left(1-(1-\theta)^{1/\nbpixel}\right)/\ln\left(1-p_1\right)$.
In the limit $p_1 \ll 1$ and for sufficiently high centiles ($\theta < 0.1$), we find that the centile of the imaging time reads: 
\begin{align} \label{eq:t0_1}
t_\theta  \underset{1 \ll \nbpixel}{\sim} \frac{\nbpixel}{\mu} \ln\left(\frac{\nbpixel}{\theta}\right) = \frac{1}{\rho \sigma} \ln\left(\frac{\volume}{\theta \sigma}\right) ,
\end{align} 
where $\mu = \nbpixel p_1$ is the mean number of observations per frame. The latter expression corresponds to the announced \refn{eq:main_message} with $r = 1$.

A key feature shared by Eqs. \ref{eq:main_message} and \ref{eq:t0_1} is the non-linear  dependence of the imaging time in terms of the number $\nbpixel$ of pixels that characterize the structure. This scaling is related to the classical coupon-collector problem \cite{Paul1961,Feller1968,Stanley1990}. The problem consists in buying a minimal number of the boxes (each containing a random coupon) in order to gather a complete collection of coupons, with a sufficiently high probability. Here, we focus on the case where each box contains, at random, either $0$ (with probability $p_0$) or $1$ coupon -- in which case the mean number of coupons per box is equal to $\mu = 1-p_0$. A straightforward proof leads to the following exact expression for the mean number of bought boxes $t$ (i.e. frames) required to collect all coupons (i.e. all pixels) is $\mathbb{E}\left[T\right] = \nbpixel (1 + 1/2 + \ldots + 1 /\nbpixel)/\mu$. If the number of coupons $\nbpixel$ is large,  the latter expression takes the asymptotic form $\mathbb{E}\left[T\right] =  \nbpixel \ln(\nbpixel)/\mu$. Adapting the identity (2) of Ref. \cite{Paul1961}, one shows that the centile of the stopping time reads $t_{\theta} = (\nbpixel/\mu) \times \ln(\nbpixel/\theta)$ in the same limit $\nbpixel \gg 1$, which corresponds to \refn{eq:Ftheta} after identification of the mean number of coupons per box to the mean number of events per frame. 

Mind that \refn{eq:t0_1} weakly depends on the risk level $\theta$, which is another characteristic property of the coupon-collector problem \cite{Paul1961,Feller1968,Stanley1990}.

Furthermore, the expression for the image completion time in \refn{eq:t0_1} only depends on the number of pixels but not their spatial organization, e.g. on the 1D, 2D or 3D nature of the structure. This is expected since pixels are considered to be independent. 


\subsection{The coupon-collector scaling holds for a near complete coverage} \label{sec:near_complete}

A simple argument shows that \refn{eq:main_message} holds even in the near total coverage, ie. when the final image should contain a significant fraction (e.g. $90\%$) of the total number of pixels within the ROI. Consider that a single pixel $i$ is missing after $t$ frames. The additional number of frames $\Delta t$ that is required to find the missing pixel $i$ is of the order of the total number of pixels, ie. $\Delta t \propto F$. This increment is small compared to the total completion time $T = T_{\mathrm{near-complete}} + \Delta t \approx F \ln(F)$. Provided that the missing fraction of pixels is small, the \textit{near-completion} time $T_{\mathrm{near-complete}}$ is approximatively equal to the completion time $T$.

\subsection{The coupon-collector scaling holds when redundant observations per pixel are required} \label{sec:redundant}

To distinguish relevant observations from spurious ones, we consider that a pixel should collect a minimal number of observations denoted $r$ to be considered as being part of the region of interest. Assuming that all pixels within the ROI are equivalent, the probability that all pixels have collected at least $r$ observations can be expressed in terms of the probability that the pixel $1$ have collected at least $r$ observations as $P \left[ \estpixel = \nbpixel \right]  = \left(  \mathbb{P}(M^{(t_\theta)}_{1} \geq r) \right)^{F}$.
We show in the SI \ref{sec:one_per_frame} \cite{SI} that, in this case, the centile of the image completion time reads
\begin{align} \label{eq:tcc2}
t_\theta \underset{1 \ll \nbpixel}{\sim} \frac{\nbpixel}{\mu} \left\lbrace \ln\left( \nbpixel/\theta \right) + (r-1) \ln\left(\ln  \nbpixel \right) \right\rbrace.
\end{align}
The latter relation corresponds to the centile of the coupon collector's problem when $r$ copies of each coupon need to be collected (see \cite{Newman1943,Paul1961}). 

We emphasize that \refn{eq:tcc2} requires that the required number of coverage $r$ is sufficiently small, ie. that $r \ll \ln(\nbpixel)$. We also generalize the result of \refn{eq:tcc2} to a multi-color imaging problem (see SI \ref{app:sec:multicolor}, \cite{SI}).



\subsection{The coupon-collector scaling holds in the presence of spatial inhomogeneities} \label{sec:spatial}

In this section, we discuss the case of a non-homogeneous rate of activation, which is particularly important in PALM (see Fig \ref{Fig2}). We model the non-homogeneity of detection events by assuming that, among pixels, the probability $p_{0}$ is distributed according to a probability distribution $\psi(q)$.  Under this assumption, the probability that there has been more than $r$ observation in a particular pixel $i$ reads: 
\begin{align} \label{eq:conditional}
\mathbb{P}(M^{(t)}_{i} \geq r) = \int^{1}_{0} \mathrm{d}q \, \mathbb{P}(M_{i} \geq r \lvert p_{0,i} = q)  \psi(q).
\end{align}
where $\mathbb{P}(M_{1} \geq r \lvert p_{0,i} = q)$ is given in \ref{eq:subset} (see also SI, \refn{eq:longtime}, \cite{SI}). From the expression in \refn{eq:conditional}, we numerically solve  the relation \refn{def:t_theta} to obtain the imaging time $t_\theta$. 

In the Experimental Comparison section \ref{sec:comparison_to_experiments}, we consider a model in which the probability $\psi$ is Gaussian distributed. We find that the coupon-collector logarithmic scaling still holds in that case. However,  the precise distribution of spatial hetereogeneities is needed in order to obtain a quantitative fit to the experimental data (see Fig. \ref{Fig5}).

\subsection{The coupon-collector scaling holds with the patch image-rendering method} \label{sec:patch}


We now consider that the image results form the accumulation of circular patches, whose radius $\sigma$ corresponds to the spatial resolution. The patch centers are distributed according to a homogeneous Poisson distribution within the region of interest, of volume $\volume$.

The study of coverage problem has a long history \cite{Flatto1973, Solomon1978}. However, analytical results concerning coverage problems in two dimensions are rather recent \cite{Zhang2004, Lan2007}. These studies were motivated by the study of the wifi coverage resulting from randomly located routers. We will make use of results concerning the expression of the centile $n_\theta$ of the number of patches required to cover a circle \cite{Flatto1973} or a square \cite{Zhang2004} by circular patches.

Here, we seek an expression of the centile time $t_\theta$, i.e. a time expressed in terms of a number of frames $t$, rather than the centile time expressed in terms of the number of patches $n$. We expect that $t_\theta = n_\theta/\mu$ where $\mu$ is the number of events per frame. Indeed, in the small patch limit $\sigma/\volume \ll 1$, full coverage events occur when the number of events is large ($n \gg 1$) in which case the number of events is simply proportional to the number of frames $t$. This approximation is further justified in the SI \cite{SI}. Therefore, following Refs. \cite{Flatto1973} and \cite{Zhang2004}, we find that \refn{eq:main_message} corresponds to the time required to obtain a $r$-fold coverage of a $D$-dimensional ROI of total volume $\volume$ by circular patches of volume $\sigma$. In particular, we obtained that $\gamma_1 = 0$ in 1D (following \cite{Flatto1973}) and that $\gamma_2 = 2$ in 2D (following \cite{Zhang2004}) and finally that $\gamma_3 = 3$. 

Remarkably, \refn{eq:main_message} takes a similar form as the coupon-collector problem from \refn{eq:tcc2}. This similarity suggests that in the limit $\sigma/\volume \ll 1$, regularly spaced patches of size $\sigma/\volume$ behave as if they were independent. Mind, however, that the expression from \refn{eq:tcc2} corresponds to a value  $\gamma_D = -1$ for any space dimension. The origin of this discrepancy at second order in the ratio $\sigma/\volume \ll 1$ is discussed in Ref. \cite{Flatto1973}. In conclusion, we have shown that both the PM and BFM lead to similar expressions for the image completion time. 


Another interest of the PM representation is that we can answer the following question: after a time $t$ has elapsed, what is the probability $P(\epsilon)$ that a hole of size $\epsilon$  in reconstructed image  corresponds to a genuine gap in the structure?
We identify $P(\epsilon)$ as being equal to the empty-space distribution defined in \cite{Solomon1978}, hence we find that
\begin{align} \label{eq:emptyspacedistribution}
P(\epsilon) = 1 - \exp(-\rho t \, \epsilon^D/\Omega),
\end{align}
where $\Omega = \pi^{D/2}/\Gamma\left[1+D/2\right]$ is the volume of a sphere of radius $1$. We expect \refn{eq:emptyspacedistribution} to hold within the BFM framework, hence providing the probability that a connected set of $N = \epsilon/\sigma$ missing pixels corresponds to a genuine hole.


\subsection{The coupon-collector scaling holds in the presence of correlations between frames} \label{sec:timecorrelation}

\begin{figure}[t!]
\includegraphics[width=7cm]{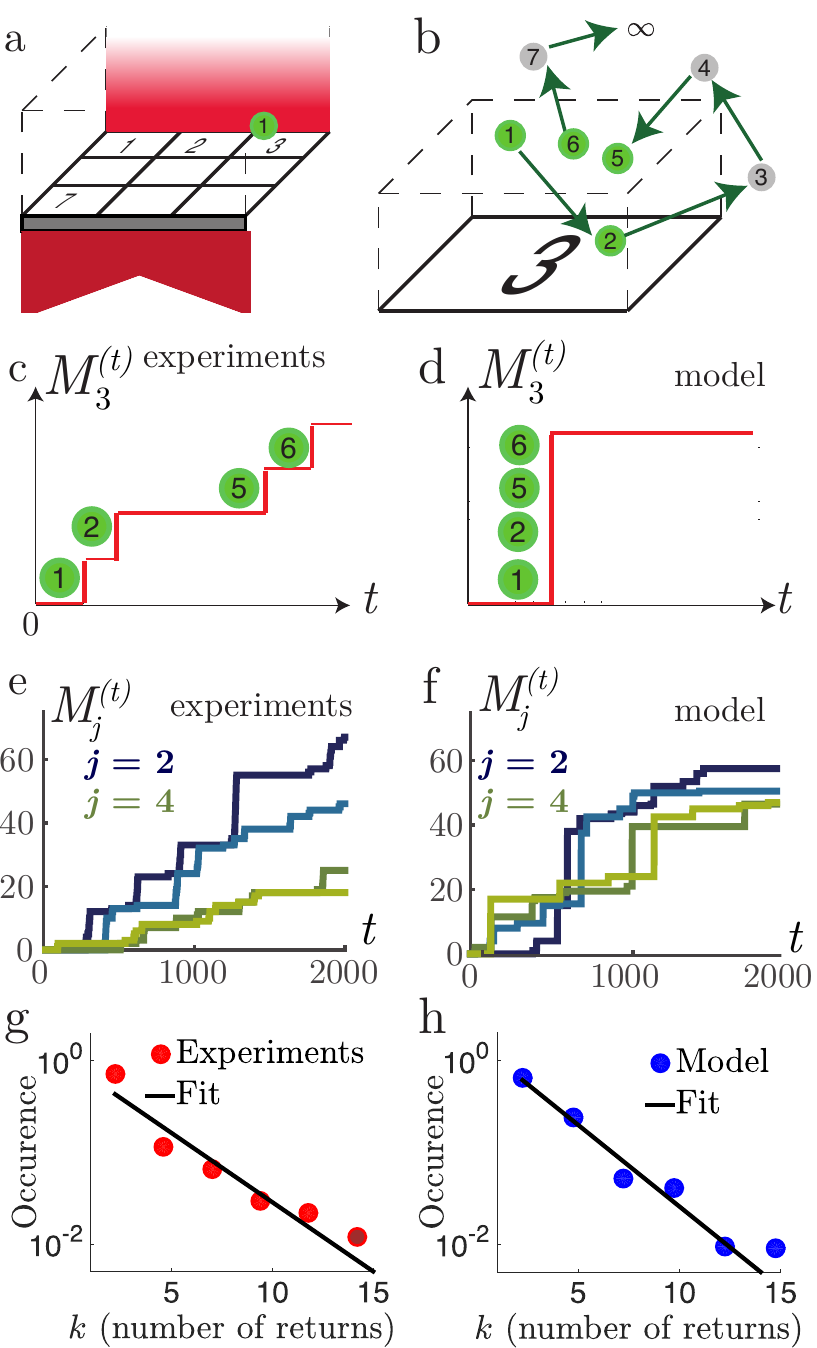} 
  \caption{(Color online) Schematic view of the stochastic Brownian TIRM technique and its modeling. (a) Global view of the total field of observation, divided into $\nbpixel$ pixels (BFM framework). (b) Zoom on pixel $3$, including a particle labelled by the frame instant.  After $t \geq 6$, the probe is not detected again. (c-d) Scheme of the evolution of the cumulative number of observations $M^{(t)}_3$ occuring in pixel $3$ (c) as seen experimentally (d) as represented in our model, where all correlated observations are collapsed into one single instantaneous events. (e-f) Evolution of the cumulative number of observations obtained, either (e) from the experimental data set of Ref \cite{Martinez-marrades2014}, or (f) from Monte-Carlo simulations, with the fitted jump distributed. The different colors correspond to different pixels ($4$ among $\nbpixel = 100$). (g-h) Distribution of jump length $p_k \propto \exp(-k/k_c)$ (in log-scale): (g) from experiments (blue circles), where the maximum likelihood estimator of the exponential model \cite{Rivoirard2012} provides the value $k_c = 2.9 \pm 0.1$ (black line); (h) from simulations, with $k_c = 2.9$ and the same number of $2792$ simulated events.}
\label{Fig3}
\end{figure}

In the Brownian scatterers experiments, the gold particle may enter, escape or return within the field of view, leading to correlated observations between successive frames. In contrast to the discussion leading to \refn{eq:Ftheta}, these temporal correlations invalidate the independence hypothesis that allows to factorize the final time probability distribution.  We encompass these correlated observations through the following box-filling model, in which the number of events per pixel and per frame is assumed to be a random variable $K$ with a general probability law $p_k = \mathbb{P}(K=k)$ for all $k \geq 0$. The statistics of $K$ encompass the effect of time-correlated observations by neglecting the time between successive correlated events. Comparison of this model to experiments is satisfactory, as visible in Figs. \ref{Fig3}(e) and (f), in which we represent the experimental data from Ref. \cite{Martinez-marrades2014} and simulated evolutions of the cumulative number of events $M^{(t)}_j$.  

We define the mean and variance of the number of observation per pixel per frame as $\nu  = \sum^{\infty}_{k=1} k p_k$ and $\sigma^2 =  \sum^{\infty}_{k=1} k^2 p_k - \nu^2$, respectively. We assume that the set of probabilities $p_k$, $k\geq 0$ is identical for each of the pixels of the structure to be imaged. In SI \ref{sec:one_per_frame} \cite{SI},  we show that the imaging completion time reads 
\begin{align} \label{eq:tcc2_general}
t_\theta \underset{\sigma \ll \volume}{\sim} (1-p_0)^{-1} \left\lbrace \ln\left(\frac{\volume}{\sigma \theta}\right) + (r-1) \ln\left[\ln\left(\frac{\volume}{\sigma}\right)\right] \right\rbrace,
\end{align}
provided that $p_1 \neq 0$. Mind that \refn{eq:tcc2_general} differs from \refn{eq:t0_1} due to the prefactor  $1 - p_0$, which is determined by the precise statistics of $K$ and may significantly differ form the value of $\mu/\volume$. In particular, at a constant total mean number of events per frame $\mu$, an increase in the mean number of correlated events $\nu$ also increases the imaging time. 

We conclude that, temporal correlations can significantly affect the value of the image completion time, yet without affecting the coupon-collector scaling of the image completion time. 



\subsection{Situations in which the coupon-collector scaling does not hold} \label{sec:coupon_not_hold}

First, when a small subset of observation is sufficient to reconstruct the image, the coupon-collector scaling should not be expected. This may include situations in which the structure can be inferred, e.g. by assuming randomly oriented linear shapes \cite{Zhang2016}. Consider that the image is considered to be complete as soon as $M \ll \nbpixel$ different pixels have been acquired. We find that the probability defined in \refn{eq:subset} is maximal after a number of frames $t_{\mathrm{opt}}(M)  \sim M/(\nu F)$ in the limit $\mu/\nbpixel \ll 1$ and $M/\nbpixel \ll 1$. Hence, the image completion time is proportional to $M$, with no logarithmic dependence on the parameters. 


Secondly, when the ratio of structure signal to the background noise is weak, a large number $r \gg \nbpixel$ of redundant observations per pixel is required; we show that the coupon collector scalings from Eqs. \ref{eq:tcc2} and \ref{eq:main_message} does not hold in this limit. Indeed, due to the central limit theorem, the number of observations collected in the pixel $j$ eventually converges with $t$ towards a Gaussian distribution:
$M^{(t)}_j \sim \quad \mathcal{N}\left(t \mu/\nbpixel,  t \Sigma^2/\nbpixel\right),$
where, as in the previous section, $\mu$ and $\Sigma^2$ are the mean and variance of the number of observation per frame within the total field of view (which, in principle, can be different from $\nu$ and $\sigma^2$ in the presence of temporally correlated noise).  Under the Gaussian assumption, we find that the probability distribution of the image completion time $T$ reads:
\begin{align} \label{eq:probagauss}
\mathbb{P}\left[ T \leq t\right] =2^{-1/\nbpixel} \left\lbrace 1 - \text{erf}\left( \frac{r-\mu  t/\nbpixel}{\sqrt{2 \Sigma^2  t/\nbpixel}}\right) \right\rbrace^\nbpixel,
\end{align}
where $\text{erf}(x) = \int^{x}_{-\infty} \! \mathrm{d}t \, \exp(-t^2)/\sqrt{\pi}$ is the error function \cite{Ryzhik1957}. In the limit of a large number of observations $r \gg \ln(\nbpixel)$, the expansion of the error function around $0$ provides the following approximate expression:
\begin{align} \label{eq:gaussian_approximate}
t_{\theta} \sim \frac{\nbpixel r}{\mu} + \sqrt{\frac{2 r \Sigma ^2}{\mu} \log \left(\frac{\nbpixel}{2 \sqrt{2 \pi } \theta }\right)},
\end{align}
in the limit $r \gg \ln(\nbpixel/\mu)$.
The key feature from \refn{eq:gaussian_approximate} is that the image completion time $t_{\theta}$ does not follow the coupon-collector scaling. Similarly to a deterministic imaging techniques, the image time scales linearly with $F$ - which should be expected since the effects of the localization randomness are all the more averaged out that the required redundancy per pixel is large.

To conclude, we have obtained analytical results for the image completion time problem in the limits $r \ll \ln(\nbpixel/\mu)$ and $r \gg \ln(\nbpixel/\mu)$ while we resorted to numerical simulation to describe the intermediate regime.

\section{COMPARISON TO EXPERIMENTS} \label{sec:comparison_to_experiments}

\subsection{Comparison to PALM experiments}

We analyzed the sequence of localization events from a silicon wafer with quasi-uniform coating. The non-uniformity in the fluorophore density leads to  heterogeneities in the value of $p_0$, i.e. the probability per frame that a pixel does not collect any observation (see Fig. \ref{Fig4}). We fit the distribution $p_0$ by a Gaussian distribution $\psi(p_0) = N \exp((p_0 - \nu_p)^2/(2 \sigma^2_p))$, where $N$ corresponds to the normalization over $p_0 \in \left[0,1\right]$.

Based on the estimation of probability distribution from \refn{eq:conditional}, we find our predicted centile time is in quantitative agreement with the analysis of  two experimental data sets. We point out that taking into account the spatial heterogeneities in $p_0$ is required to obtained the quantitative fit represented Fig. \ref{Fig4}d.

\subsection{Comparison to TIRM experiments}

\begin{figure}[t!]
  \includegraphics[width=8.60cm]{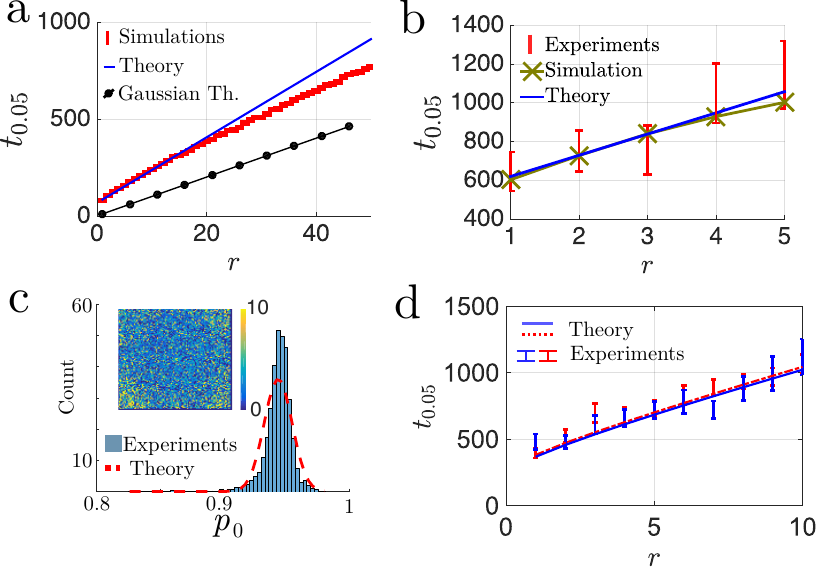} 
  \caption{(Color online) Centile $t_{0.05}$ of the image completion time as a function of the required number of redundant observations per pixel $r$. (a) Simulations with at most one observation per pixel and per frame, $\nbpixel = 15$ and $p_1 = 0.1$. The analytical expression from \refn{eq:tcc2} (solid blue line) provides a better fit of simulations (red error bars, obtained by bootstrapping \cite{Efron1979}) than the approximate solution from \refn{eq:gaussian_approximate} (black circle) . 
  (b) Comparison of our theoretical expression to the TIRM experiments from \cite{Martinez-marrades2014}, with $\nbpixel = 15$: (red error bars) centile estimation from the experiments; (blue solid line) theoretical prediction from \refn{eq:tcc2} with a  jump probability distribution $p_k \propto \exp(-k/k_c)$ with $k_c = 2.9$; (green crosses) stochastic simulations. 
  (c-d) PALM imaging of a silicon wafer with quasi-uniform coating in fluorophores (the ROI corresponds to the whole field of view, with $N = F= 100$). (c) Non-uniform fluorophore density leads to spatial heterogeneities in $p_0$, i.e. the probability per frame that a pixel does not collect any observation: the experimental distribution (blue boxes) is fitted by a Gaussian (dashed red line). Inset: field of view in terms of the number of collected observations per pixel. (d) Centile $t_{0.05}$ for (blue and red) two distinct samples with identical concentration of fluorophores: (solid lines)  analytical expression and  (error bars) experimental result.}
\label{Fig4}
\end{figure}

\begin{figure}[t!]
  \includegraphics[width=8.60cm]{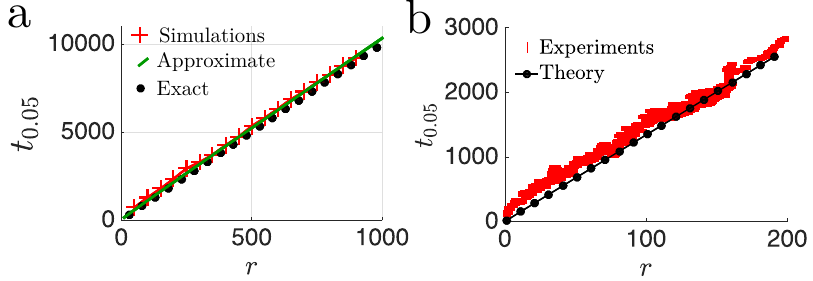} 
  \caption{(Color online) Centile $t_\theta$ of the image completion time as a function of the required number of events per pixel in the regime $r \geq \ln(\nbpixel)$. 
(a) Simulations with at most one observation per pixel, $\nbpixel = 15$ and $p_1 = 0.1$: (red cross) centile from stochastic simulations; (solid green line) approximate solution from \refn{eq:gaussian_approximate}; (blue circles) exact centile time obtained by numerical inversion of \refn{eq:probagauss}.
(b) TIRM experiments from Ref. \cite{Martinez-marrades2014}, with $\nbpixel = 4$: (red bars) centile estimation from experiments, where the error is estimated by bootstrapping \cite{Efron1979}; (black dots) theoretical prediction from \refn{eq:gaussian_approximate}.}
\label{Fig5}
\end{figure}

We represent the TIRM experiments data from Ref. \cite{Martinez-marrades2014} within the BFM framework and we include temporal correlations between frames. First, the mean number of particles per frame and over the whole field of view reads $\mu = 0.70$. Secondly, the jump distribution is estimated as follow: two successive events are assumed to correspond to the return of the same particle if (i) they occur within the same pixel and (ii) they are separated by a time interval of less than $\Delta = 5$ frames. The experimental histogram is fitted by the distribution $p_k = A_{k_c} \exp(-k/k_c)$, where $A_{k_c} = 1/(1-\exp(-1/k_c))$, and $k_c = 2.9$ (see Fig. \ref{Fig3}.g.).  This leads to a mean jump size $\nu= 1/(1-\exp(-1/k_c)) = 3.4$ and a variance $\sigma^2 = 1/(\cosh(1/k_c) - 1) = 17$. We check that our results weakly depend on the specific value attributed to the separation time $\Delta$.

As described in Figs. \ref{Fig4}a and \ref{Fig5}b, we show that our theoretical expressions from Eqs. (\ref{eq:tcc2_general}) and (\ref{eq:gaussian_approximate}) both fit to the experimental estimation of the centile time in their respective validity range. We point out that a straightforward implementation of \refn{eq:tcc2}, which would neglect temporal correlations, leads to a value that is an order of magnitude lower than what is experimentally observed.

In the SI \cite{SI}, Sec. \ref{sec:app:justification}, we justify that a large number of redundant observations $r \approx 4\cdot 10^{3}$ is required in order to obtain a reliable measure of the penetration length $\beta$.

\section{REAL-TIME ESTIMATION OF THE RISK OF STOCHASTIC ABERRATION} \label{sec:estimator_requiredimagingtime}

\begin{figure}[t!]
\includegraphics[height=8.0cm]{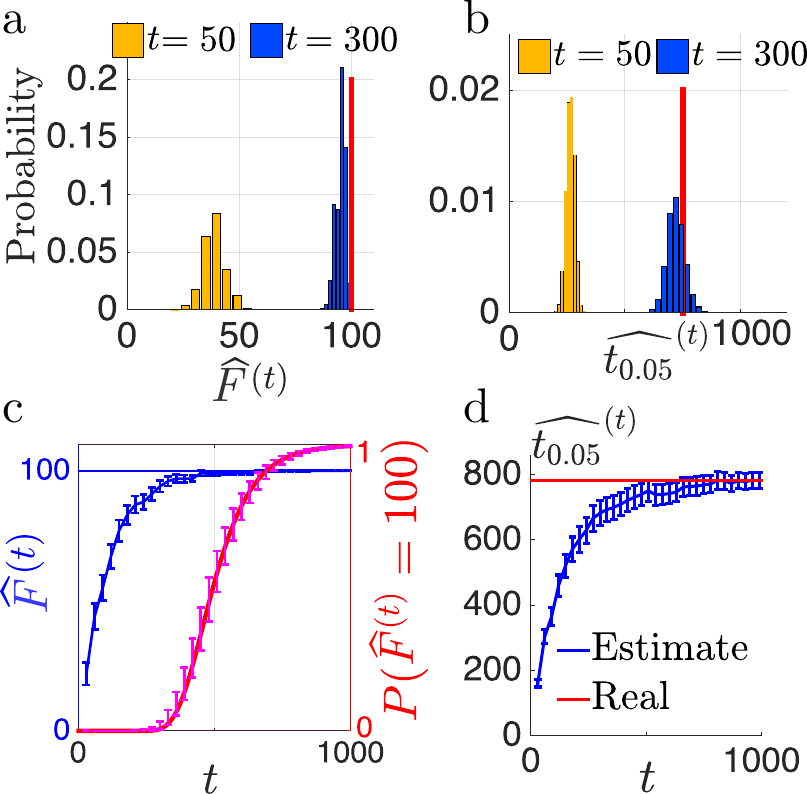}
  \caption{(Color online) Numerical simulation of the real-time imaging method (Sec. \ref{sec:estimator_requiredimagingtime}), where the total number of pixels is $\nbpixel = 100$, the probability of an event observation per frame and per pixel is $p_1 = 10^{-2}$, averaging is performed over $10^4$ samples. 
   (a) Histogram of the values of the estimator $\estpixel$ after a number $t = 50$ of observations (light orange) and  $t = 300$ (dark blue); (red vertical line) the limit value is $\nbpixel = 100$. 
  (b) Histogram of the values of the estimator  $\widehat{t_{0.05}}^{(t)}$  after a number $t = 50$ of observations (light orange) and  $t = 300$; (red vertical line) the limit value is $t_{0.95}= 760$. 
  (c) Evolution of the mean value of the estimator $\estpixel$ (solid blue curve), together. We also represent the probability  $P(\estpixel=\nbpixel)$ for the image completion (solid magenta curve). In both cases, error bars indicate the standard deviation estimated from the random sampling. Hence a single random realization (i.e. a single experiment) is sufficient to obtain a good estimate of $\estpixel$ and $P(\estpixel=\nbpixel)$.
  (d) Probability distribution for the estimator $\widehat{t_{0.05}}^{(t)}$. The distribution converges to the centile $t_{0.95}= 760$ (red vertical line) as $t$ increases. The value $\widehat{t_{0.95}}^{(50)} = 264$ is significantly larger than the current number of frame $t = 50$: this is consistent with the conclusion that more observations are required. }
\label{Fig6}
\end{figure}

Experimentally, the two quantities  $\nbpixel$ and $\mu$ are unknown {\it a priori}. These quantities are indeed associated to the structure to be imaged, whose properties are unknown prior to imaging. Here, we propose a real-time procedure to determine whether we can safely consider that the image is complete.  We emphasize that this procedure is not specific to a choice of image representation method, nor on the required number of redundant observations $r$. 

We evaluate the probability that the image is complete based on the two estimators $\estpixel$ and $\widehat{\mu}^{(t)}$ of number of pixels and of the mean number of events per frame, respectively.  For example, in the BFM with $r = 1$, the estimator of the image completion probability reads:
\begin{align}  \label{eq:estimateofprobability}
\widehat{\mathbb{P}}\left[ \estpixel  = \nbpixel \right] = \left(1-(1- \widehat{\mu}^{(t)}/\estpixel)^{t}\right)^{\estpixel}.
\end{align} 
We represent the evolution of the estimated probability corresponding to \refn{eq:estimateofprobability} in Fig. \ref{Fig6}c. We set the values to $\nbpixel = 100$ and $p_1 = 10^{-2}$. At $t = 300$ the image completion probability is lower than  $4 . 10^{-3}$: hence more frames are needed. The question is now to determine how many additional number of frames are required.

Our analytical expression of the image completion time can then be used to infer the required additional number of frames. For example, the quantity $\widehat{t_\theta}^{(t)} = (\estpixel/\widehat{\mu}^{(t)}) \ln(\estpixel/\theta)$ is an estimator of the image completion time, where we consider the BFM with $r=1$ for simplicity. We represent the convergence of the latter estimator to the expected value of the centile time $t_\theta$ in Fig. \ref{Fig6}b,d. After $t= 300$ frames, we estimate that about $420$ additional frames are required, which is consistent with the theoretical value of the centile time $t_\theta = 760$.

Based on the estimators $\widehat{\mathbb{P}} [ \estpixel  = \nbpixel ]$ and $\widehat{t_\theta}^{(t)}$, we propose the following  procedure to analyse an imaging experiment in which $t$ frames have been collected:
\begin{enumerate}
\item Compute the estimators of the number of pixels ($\estpixel$) and of the mean number of events per frame ($\widehat{\mu}^{(t)}$).
\item Compute the estimator of the probability that the image is complete. If this estimator is higher than a desired confidence threshold, the imaging process can be stopped.
\item Otherwise, compute the estimated image completion time $\widehat{t_\theta}^{(t)}$. Perform $\widehat{t_\theta}^{(t)} - t$ additional frames and return to step $1$ with the substitution $t \leftarrow \widehat{t_\theta}^{(t)}$.
\end{enumerate}

The above procedure is not specific to any particular criteria for the image completion. For example, if a large redundancy is required ($r \gg \ln(\nbpixel)$), one should use the expressions of \refn{eq:probagauss} for the probability that the image is complete and \refn{eq:gaussian_approximate} for the image completion time. In the SI \ref{sec:app:area} \cite{SI}, we provide an expression for the probability that the image is complete within the PM framework.

\vskip1cm

\paragraph*{Conclusion}

Our theoretical model provides a unified framework to describe the temporal resolution of several types of stochastic microscopy techniques. These include PALM, in which a large number of fluorescent probes are attached to the sample and are stochastically activated, or techniques in which a smaller number of scattering probes stochastically explore the imaged region. We derive analytical expressions for the centile of the imaging time for several types of image completion criteria. When a sufficiently low number of accumulated events per pixel are required, the temporal resolution is shown to be logarithmically coupled to the spatial resolution (pixel size), due to the spatial redundancy of detection  events. However, the temporal resolution becomes linearly coupled to the spatial resolution when a large spatial redundancy of events is needed, as the effects of the localization randomness are averaged out. Our results on the imaging time are readily applicable to estimate the minimal time required to reliably characterize spatial patterns by stochastic imaging, with applications ranging from the detection of protein clusters by PALM \cite{Changede2015a} to the detection of the electromagnetic field around nano-antennas by Brownian particles \cite{Martinez-marrades2014}.

\vskip1cm

\paragraph*{Supplementary material} Electronic supplementary material is available at XXX or via YYY.

\paragraph*{Authors’ contributions}
A. M. M. and G. T. carried out the gold nano-particles experiments and localization analysis, and instigated the theoretical problem. R. C. carried out the PALM experiments and localization analysis. J.-F. R. performed the theoretical calculations, simulations, centile time analysis of the experiments and wrote the manuscript. We have no competing interests.

\paragraph*{Acknowledgements}
We thank Xu Xiaochun (MBI Microscopy core) for designing the localization code of the PALM setup, V. Studer, M. Coppey and B. Hajj for enlightening  discussion on the PALM technique, and S. Tlili for comments on the manuscript.

%

\appendix 

\maketitle

\section{Description of the PALM experiments}

\subsection{Materials and methods}

\paragraph*{Silane functionalization} Biotinylated silane (Methoxy Silane PEG biotin) was mixed with Methoxy Silane in concentration equivalent to have a final concentration of $10^3 \, \mathrm{molecules}/\mu \mathrm{m}^2$ of Biotin on the surface. Clean cover slips were coated with this silane mixture using vaporization under vacuum. Silane functionalized cover slips were then washed with PBS and incubated with Dylite650 Neutravidin for $1$ hour followed by a subsequent wash before PALM imaging. 

\paragraph*{Cell Culture and sample preparation} Mouse embryonic fibroblasts were grown in DMEM media containing 1mM sodium pyruvate and $10\%$ fetal bovine serum at $37^{\circ} \ C$ with $5\%$ of carbon dioxyde. Cell were spread for $4$ to $6$ hours on a fibronectin coated glass dish. Spread fibroblasts were then were fixed with $4\%$ formaldehyde at $37^{\circ} \ C$ for $10$ minutes followed by mild detergent permeablization with $0.5\%$ Triton X 100 for 15minutes. Alexa647- Phalloidin was used to label actin ($125 \mathrm{nM}$ for $1$ hour immediately prior to imaging).  Multiflurophore beads of $0.14 \, \mu m$  diameter (Spherotech, Cat no. FP0257-2) were added as fiducial markers for PALM imaging. 

\subsection{Photoactivated light microscopy (PALM)}

Fresh imaging buffer was made for every sample. Imaging buffer contained we oxygen scavenging imaging buffer constituting of the following solutions with a volume ratio of 90:10:1.
(1) 50 mM Tris-HCl (pH 8.0), 10 mM NaCl, 10
(2) 1M mercaptoethylamine with pH adjusted to 8.5 using HCl
(3) Anti-bleaching oxygen scavenger system containing $14 \, \mathrm{mg}$ Glucose Oxidase, 
50 µl Catalase (17 mg/ml) in $200 \, \mu L$ 10mM Tris-HCl (pH 8.0), and $50 \, \mathrm{mM}$ NaCl DPBS solution. 
The samples were sealed with parafilm to prevent air exchange.

PALM imaging was performed using Ziess Elyra. 100X objective (Alpha Plan Apochromat 100X oil NA 1.46,) with 1.6 magnification to have a final pixel size of $100 \, \mathrm{nm}$ by $100 \, \mathrm{nm}$ was used. The camera on the system is Andor iXon DU897 $512x512$ electron multiplier CCD camera. A total of $2 \cdot 10^{4}$ images were collected with continuous streaming at $50 \, \mathrm{ms}$ per frame for each sample. PALM images were reconstructed using a custom-made maximum likelihood software \cite{Changede2015a}. 

\section{Image completion time within the box-filling framework}

\subsection{Maximum likelihood estimator of the number of pixel in the PALM}
\label{sec:app:Festimator}

We consider the result of a particular simulation or experiment in which the cumulative number of observation within the pixel $j$ reads: $M^{(t)}_j = k_j$ for all $j \leq \nbpixel$. The likelihood of such outcome is defined as the probability:
\begin{align} \label{eq:likelihood}
\mathbb{P}\left(M^{(t)} = k \right) = \prod^{\nbpixel}_{j=1} \binom{t}{k_j} p^{k_j}_1 (1-p_1)^{t-k_j} 1_{\estpixel \leq \nbpixel},
\end{align}
for all $k \geq 0$, and where $1_{\estpixel \leq \nbpixel}$ is the indicative function, equal to 1 if $\estpixel \leq \nbpixel$ and $0$ otherwise. The product in \refn{eq:likelihood} spans from $j = 1$ to $j = \nbpixel$ as $M^{(t)}_j = 0$ with probability $1$  for all pixels which do not correspond to the structure of interest. Due to the indicative function, the global minimum of \refn{eq:likelihood} is achieved for $\nbpixel = \estpixel$ -- therefore $\estpixel$ is called the maximum likelihood estimator of $\nbpixel$. 

\subsection{Proofs for the coupon-collector scaling in the regime $1 < r < \ln(\nbpixel)$}

\subsubsection{One observation per pixel per frame}  \label{sec:one_per_frame}

Here we consider the case where the number of observations of a pixel at each frame is either $0$ or $1$ (i.e. $p_k = 0$ for all $k > 1$). \\

We define the probability $q_{j}^{(t)} $ that the pixel $m$ has been observed a number $j$ times at the time $t$: $q_{j}^{(t)} = \mathbb{P}(M^{(t)}_m = j)$. Successive observations are considered as independent in time, hence $q^{(t+1)}_{j} = p_0 q^{(t)}_{j} + p_1 \, q^{(t)}_{j-1}, \quad 1 \leq j \leq r-1$ for all $1 \leq j < r$. As we are interested in the time required to reach the state $j = r$, we consider the state $j = r$ to be an absorbing state $q^{(t+1)}_{r} = q^{(t)}_{r} + p_1 \, q^{(t)}_{r-1}$. As soon as $j \leq t$, the probability to have reached $j \leq r-1$ observations of the pixel is:
\begin{align}
\mathbb{P}(M^{(t)}_m = j) =  \frac{t!}{(t-j)! \, j!} \, p_1^{j} \, (1-p_1)^{t-j} , \qquad j \leq r-1,
\end{align}
from which we deduce the probability that the pixel has been observed at least $r \geq 2$ times is: $q^{(t)}_{r} =  1 - \sum^{r-1}_{j=0} q^{(t)}_{j}$. In the long-time limit $1 \ll t$, $t!/(t-j)! \sim t^{j}$ and the absorption probability $q^{(t)}_{r}$ tends to $1$ as
\begin{align} \label{eq:longtime}
\mathbb{P}(M^{(t)}_m = r) = 1 - t^{r-1} \frac{p_0^{t}}{(r-1)!} \left(\frac{p_1}{p_0}\right)^{r-1} \quad \mathrm{for} \quad  1 \ll t.
\end{align}

The probability that all pixels have been observed $r$ times at the time $t$ is $(q_{r}^{(t)})^{\nbpixel}$.
We are interested in the centile time $t_\theta$ given by the condition: $P\left(\left\lbrace \nbpixel \delta_{jr} \right\rbrace_{j}\right)  =  (q_{r}^{(t_\theta)})^{\nbpixel} = 1-\theta$. In order to obtain a simple explicit expression for $t_\theta$, we approximate the probability $q_{r}^{(t)}$ by its long-time behavior from \refn{eq:longtime} (which is valid for $\theta$ is sufficiently small or for $\nbpixel$ sufficiently large) to obtain that:
\begin{align} \label{eq:tcc}
1-(1-\theta)^{1/\nbpixel} =  \frac{p_0^{t_\theta}}{(r-1)!} \left(t_\theta \frac{p_1}{p_0}\right)^{r-1}.
\end{align}
Given that $1-(1-\theta)^{1/\nbpixel}  \sim \theta/\nbpixel$ in the limit $\theta \ll 1$, we obtain from \refn{eq:tcc}: 
\begin{align} \label{eq:tap}
\ln(p_0)  t_\theta + (r-1) \ln\left( \frac{p_1} {p_0} t_\theta  \right) =  \ln\left(  \frac{\theta}{\nbpixel} \right) + \ln\left[ (r-1)! \right],
\end{align}
which, in the limit  $r \ll \nbpixel$, leads to:
\begin{align} \label{eq:tcc2_general_2}
t_\theta = - \frac{\left\lbrace \ln\left(\frac{\nbpixel}{\theta}\right) + (r-1) \ln\left[\frac{p_1}{p_0(- \ln(p_0))} \ln\left(\frac{\nbpixel}{\theta}\right)\right] + C_1 \right\rbrace}{\ln(p_0)},
\end{align}
where $C_1$ is a constant of $\nbpixel$. 
In the regime of rare hits ($1 - p_0 \ll 1$), then $\ln(p_0) = \ln(1-(1-p_0)) = -(1-p_0) = - p_1 = - \mu/\nbpixel$, where $\mu$ is the mean number of hits per frame, \refn{eq:tcc2_general_2} then reads
\begin{align} \label{eq:tcc_N}
t_\theta \underset{1 \ll \nbpixel}{\sim} \frac{\nbpixel}{\mu} \left\lbrace \ln(\nbpixel/\theta) + (r-1) \ln(\ln(\nbpixel/\theta))  + C_1/\nbpixel \right\rbrace,
\end{align}
in the limit $r \ll \nbpixel$. This proves the relation of \refn{eq:tcc2}.

\subsubsection{Random number of observations per frame} \label{sec:random_number}

In this section, we consider that, at each frame, the number of observations of a given pixel is random variable equal to (i) $0$ with probability $p_0$ and (ii) to a value $k \in \left[0, r\right]$ with a probability law $p_k$. 

Following the method of the previous paragraph \ref{sec:one_per_frame}, we consider the coverage dynamic for a single pixel. The probability that the single pixel has been observed $j$-times, with $1 \leq j \leq r-1$, during a sequence of $t$ frames is:
\begin{align} \label{eq:qwithjumps}
q^{(t)}_{j} &= \sum \frac{t!}{(t - j_u)! \ldots j_r!} p_{0}^{t} \left( \frac{ p_{1}}{ p_{0}} \right)^{j_1} \ldots \left( \frac{ p_{r}}{ p_{0}} \right)^{j_r}, 
\end{align}
where (i) the sum holds over the sets of indices $(j_1, \ldots j_r)$ that guarantee the condition that $\sum^{r}_{m=1} m j_m = j$, and (ii) $j_u = \sum^{r}_{m = 1} j_m$ is the total number of adsorption events.

The imaging time $t_\theta$ is defined by the equation: $(q_{r}^{(t_\theta)})^{\nbpixel} = 1-\theta$, in agreement with \refn{def:t_theta}. In order to obtain a more explicit expression for $t_\theta$, we focus on two simple cases where the asymptotic behavior of $q^{(t)}_{r} =  1 - \sum^{r-1}_{j=0} q^{(t)}_{j}$ for $t \gg 1$ can be analytically studied.

We first review the case where steps are all of equal height: $p_k = p_s \delta(s-k)$ (e.g. $p_1 = 0$) and $r$ is a multiple of $s$ i.e. there exists $q$ such that $r = qs $. Then the situations amounts to the case considered in the section \ref{sec:one_per_frame}, with the substitution $r \leftarrow q$ and  $p_1 \leftarrow p_s$.

The second case relies on the hypothesis that $p_1 > 0$. The set of indexes that maximizes the exponent $j_u$ in  \refn{eq:qwithjumps} under the constraint that $\sum^{r}_{m = 1} m j_m = j$ is $(j, 0, \ldots 0)$. Moreover,  $j=r-1$ maximizes the exponent $j_u = j_1 = j$. At the leading order in $t$, \refn{eq:qwithjumps} reads
\begin{align} \label{eq:longtime2}
\mathbb{P}(M^{(t)}_m \geq r) = 1-  t^{r-1} \frac{p_0^{t}}{(r-1)!} \left(\frac{p_1}{p_0}\right)^{r-1},
\end{align}
which is identical to \refn{eq:longtime}, and leads to the scaling \refn{eq:tcc_N}, and which therefore proves \refn{eq:tcc2_general}.

\subsection{Multi-colored images} \label{app:sec:multicolor}
Our results are readily adaptable to the case of a colored image, i.e. resulting from the combination of several channels of light emission produced by different imaging probes.  This technique is frequently used in cell biology to image simultaneously actin, myosin and other proteins  \cite{Shroff2007}.  The number of distinct types of imaging probes is denoted $C$ ; the number of pixels that contain the $j$--type probe is denoted $\nbpixel_{j}$ ; the probability (per pixel and per frame) to detect an imaging probe is denoted $p_{1,j}$. The estimators $\estpixel_j$ are defined similarly to \refn{eq:Festimator}. The imaging time is now defined by the relation $\mathbb{P} \left( \left\lbrace \estpixel_1  = \nbpixel_1 \right\rbrace \cap \ldots \cap  \left\lbrace \estpixel_C  = \nbpixel_C \right\rbrace \right) = 1 - \theta$. Assuming that the imaging probes act independently, the imaging time $t_\theta$ (with $C$ colors) is given by the relation:
\begin{align} \label{eq:colored}
\prod^{C}_{j=1} \left(1-(1- p_{1,j})^{t_\theta}\right)^{\nbpixel_j} = 1 - \theta.
\end{align} 
The imaging time defined in \refn{eq:colored} exhibit a coupon collector scaling in the following two situations:

(i) if the emission probabilities are identical for all probes (i.e. $p_{1,j} = p_1$), the expression from \refn{eq:t0_1} holds after the substitution of (a) $\nbpixel$ by $\nbpixel_1 + \ldots + \nbpixel_C$, and (b) $\mu$ by $\mu = \mu_1 + \ldots +  \mu_C$, which corresponds to the total number of events per frame. Therefore, if we further assume that $\nbpixel_j = \nbpixel$ for all $j$, we show that the imaging time exhibits a coupon-collector type behavior ($C \nbpixel \ln(C \nbpixel)$) in terms of total number of pixels $C \nbpixel$.

(ii) if one channel is characterized by a weak blinking probability compared to all the other probes (e.g. $p_{1,1} \ll p_{1,j}$ for all other $j$, then it will likely be the limiting factor in the imaging process, in which case \refn{eq:t0_1} holds after the substitution of $\nbpixel$ and $p_1$ by $\nbpixel_1$ and $p_{1,1}$. 

\subsection{Inhomogeneous field of view}

We first consider the case in which at most one observation per pixel and per frame can occur. We suppose that the probability for $p_0$ in each pixels follows a Gaussian distribution, with:
\begin{align}
\mathbb{P}(p_{0,i} = q) = \frac{\exp\left(-(q - \hat{p}_0)/(2 \sigma^2)\right)}{\int^{1}_{0} \mathrm{d}u \exp\left(-(u - \hat{p}_0)/(2 \sigma^2)\right)}.
\end{align}
We express the conditional relation that
\begin{align} \label{eq:conditional2}
\mathbb{P}(M^{(t)}_{i} \geq r) = \int^{1}_{0} \mathrm{d}q \, \mathbb{P}(M_{i} \geq r \lvert p_{0,i} = q)  \mathbb{P}(p_{0,i} = q).
\end{align}
We notice that $\mathbb{P}(M_{1} \geq r \lvert p_{0,i} = q)$ is given by \refn{eq:longtime}. The imaging time is now defined by the relation $\mathbb{P} \left( \left\lbrace M^{(t)}_{1} \geq r \right\rbrace \cap \ldots \cap  \left\lbrace M^{(t)}_{F} \geq r \right\rbrace \right) = 1 - \theta$. Assuming that each pixels are independent and using \refn{eq:conditional}, the imaging time $t_\theta$ is defined by the relation 
\begin{align} \label{eq:inhomogeneous}
\left( \int^{1}_{0} \mathrm{d}q \, \mathbb{P}(M^{(t)}_{1} \geq r \lvert p_{0,i} = q)  \mathbb{P}(p_{0,1} = q) \right)^{F} = 1 - \theta.
\end{align} 

From the experimental data set, we estimate the local value of $p_{0,i}$ for each pixel $i$. For a number of pixel $F = 2500$, we find that the values of $p_{0,i}$ can be qualitatively considered as Gaussian distributed (see Fig. ). The mean probability is $E(p_0) = 0.95$ and the standard deviation is $\sigma_s = 0.010$. We point out that the latter value of $\sigma_s$ is significantly larger than the standard deviation $\sigma_u < 10^{-5}$  associated to the uncertainty in the estimation of $p_{0,i}$ in each pixels.


\section{Patch-method} \label{sec:app:arcs}

In the following sections \secn{eq:evolution_mea_coverage} to \secn{eq:evolution_mea_coverage3}, we focus on the 1D--coverage problem of a circle by circular arcs.

\subsection{Evolution of the mean coverage} \label{eq:evolution_mea_coverage}

We denote by $\widehat{Y}^{(t)}$ the fraction of points which are still left uncovered after $t$ frames.  For a single patch, $\nbpixel = 1$, the mean uncovered area is $\mathbb{E}\left[\widehat{Y} \lvert 1 \, \mathrm{events}\right] = 1 - \sigma/\volume$. Since patches occur at independent positions within $\volume$, we can factorize the mean covered area after $n$ number of patches: $\mathbb{E}\left[\widehat{Y} \lvert n \, \mathrm{events}\right] = (1-s/\volume)^n$. After averaging over the distribution of the number of events up to the time $t$, we find that the mean covered volume after $t$ frames reads \cite{Solomon1978}
\begin{align}
\mathbb{E}\left[\estvolu/\volume\right] = 1 - \exp(- \mu t \sigma/\volume),
\end{align}
which holds for arbitrary values of $\sigma/\volume$ and $t$. 

\subsection{Evolution of the probability distribution as a function of the number of frames} \label{eq:evolution_mea_coverage3}
Following \cite{Solomon1978}, we express the exact probability that $n$ patches cover the whole circle as
\begin{align} \label{eq:no_gaps_after_n_events}
\mathbb{P}_{\volume}\left[ 0 \, \mathrm{gap} \lvert n \, \mathrm{events}\right] =  \sum^{k}_{j=0} {n \choose j} (-1)^j \left(1 - j \frac{\sigma}{\volume} \right)^{n-1},
\end{align}
where $k$ is the greatest integer smaller than $\volume/\sigma$, and $\mathbb{P}_{\volume}$ denotes the probability measure with a structure volume equal to $\volume$. 

The probability that no gap remains after a time $t$ then reads: $P\left[ 0 \, \mathrm{gaps} \right] = \sum^{\infty}_{n = k+1} P\left[ 0 \, \mathrm{gaps} \lvert n \, \mathrm{events}\right] P\left[  N^{(t)}_e = n \right]$. We define the set of coefficients that
\begin{align*}
a^{(t)}_j &= \sum _{n=k+1}^{\infty } \frac{\exp (-\mu t) \left(\binom{n}{j} ((\mu t) (1- j \sigma/\volume))^n\right)}{n!}  \\ &= (\mu t)^j e^{- j (\mu t) \sigma/\volume} (1 - j \sigma/\volume )^j \gamma_j,
\end{align*}
where 
\begin{align*}
\gamma_j = \frac{\Gamma (-j+k+1)-\Gamma (-j+k+1,(\mu t) (1 - j \sigma/\volume ))}{\Gamma (j+1) \Gamma (-j+k+1)}.
\end{align*}
With these definitions, we conclude using Fubini's theorem that the probability reads
\begin{align} \label{eq:no_gaps_after_t_frames}
P\left[ T \leq t \right] = P\left[ 0 \, \mathrm{gap} \right] = \sum^{k}_{j=0} (-1)^j \frac{a^{(t)}_j }{1 - j \sigma/\volume}.
\end{align}
We mention that, from Ref. \cite{Solomon1978}, the probability that a number $i \leq k$ of gaps remains after $n$ events, denoted $P\left[ i \, \mathrm{gaps} \lvert n \, \mathrm{events} \right]$, reads:
\begin{align}
& {n \choose i}  \sum^{k-i}_{j=0} {n-i \choose j} (-1)^j \left(1-(i+j) \frac{\sigma}{\volume}\right)^{t-1},
\end{align}
which is the continuous analogue of \refn{eq:subset} in the box-filling model. 

\begin{figure}[t!]
\includegraphics[width=8.6cm]{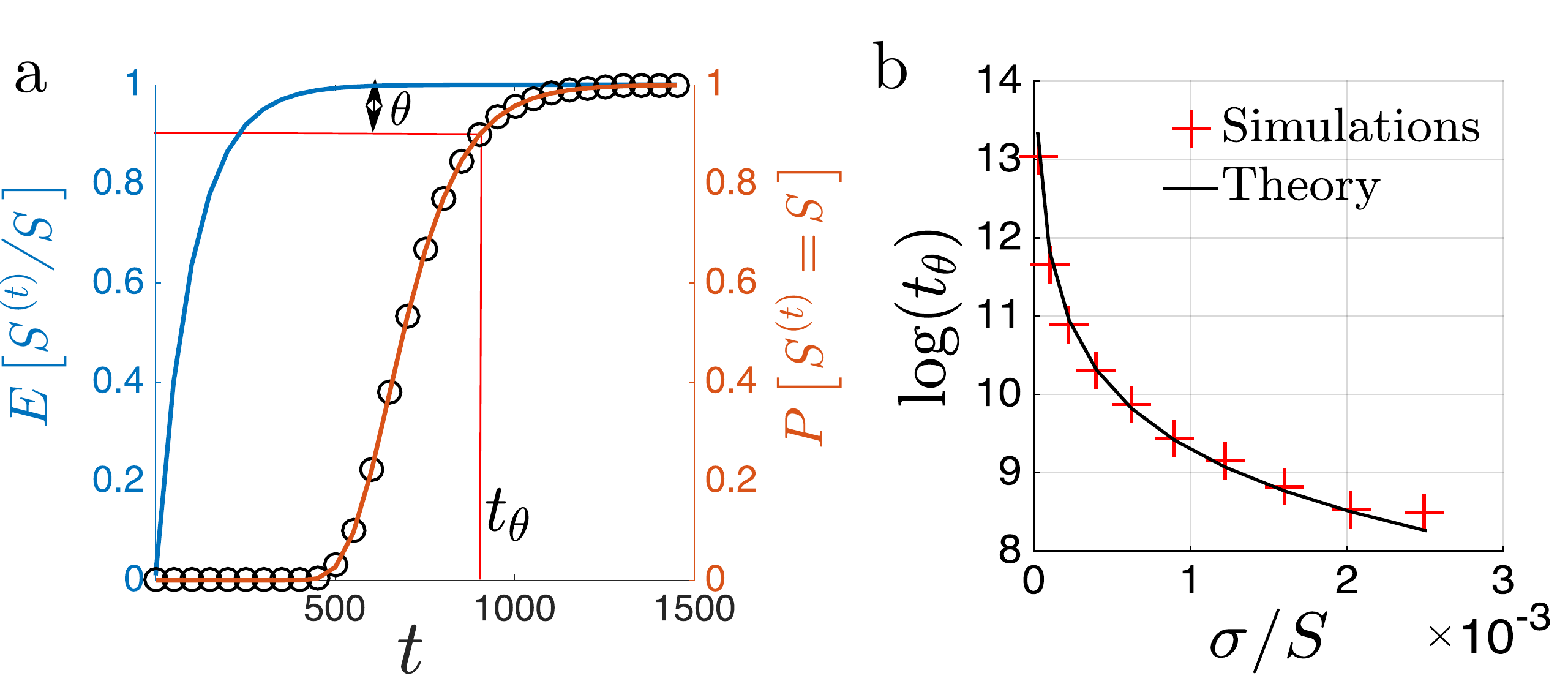}
  \caption{(Color online) (a) Probability of the full coverage of a circle of circumference $\volume$ by a number of $t$ of patches (arcs segments) of size $\sigma/\volume = 10^{-2}$: (blue solid line) mean covered area $\widehat{Y}^{(t)}$ after $t$-frames (with $1$ event per frame) ; (orange line) exact probability of a full coverage at time $t$ from \refn{eq:no_gaps_after_n_events}
 (black circles) estimated probability for a full coverage at time $t$, as constructed from the mean covered area $\widehat{Y}^{(t)}$, from \refn{eq:estimator_of_the_full_coverage}. (b) Evolution of the centile of the image time as a function of the ratio of the volume covered at each localization event ($\sigma$) by the total volume of the field of view ($S$).}
\label{fig_si_1}
\end{figure}

\subsection{The area estimation problem} \label{sec:app:area}

Following the method presented in the final section in the main text, we propose an estimator of the risk that the image is not complete, based on the collected information after $t$ frames. In particular, we provide an estimate of the additional number of frames that should be taken to obtain a complete image at a given confidence ($\widehat{t}^{(t)}_\theta$). This estimator rely on the current covered area of the experimental realization $\estvolu$, which corresponds to
a maximum-likelihood estimator of the volume of the structure to be imaged. We assume that the covered length $\estpixel$ is well described by the mean covered area at time $t$, hence that 
\begin{align} \label{eq:estimator_of_the_full_coverage}
\widehat{\mathbb{P}}^{(t)}\left( \estvolu = \volume \right) \approx \mathbb{P}_{\volume\left(1 -e^{-\sigma t/\volume}\right)}\left[ 0 \, \mathrm{gap} \right] ,
\end{align}
where $\mathbb{P}_{\volume\left(1 -e^{-\mu \sigma t/\volume}\right)}$ refers to the probability defined in \refn{eq:no_gaps_after_n_events}, that the total area to be covered is $\volume\left(1 -e^{-\mu \sigma t/\volume}\right)$.
We test the method on Fig.  SI. \ref{fig_si_1}a, in a case where the ratio $\volume/\sigma$ is known, and we find that the approximation of \refn{eq:estimator_of_the_full_coverage} is very satisfactory. Following the procedure defined in \secn{sec:estimator_requiredimagingtime}, we estimate the additional number of frames required to obtained a complete image, which can be deduced from \refn{eq:main_message}, through the substitution of $\volume$ and $\mu$ by their corresponding estimators.

\section{Calculation of the minimum number $r$ of observations per pixel} \label{sec:app:justification}

\begin{figure}[t!]
	{\includegraphics[width=8.25cm]{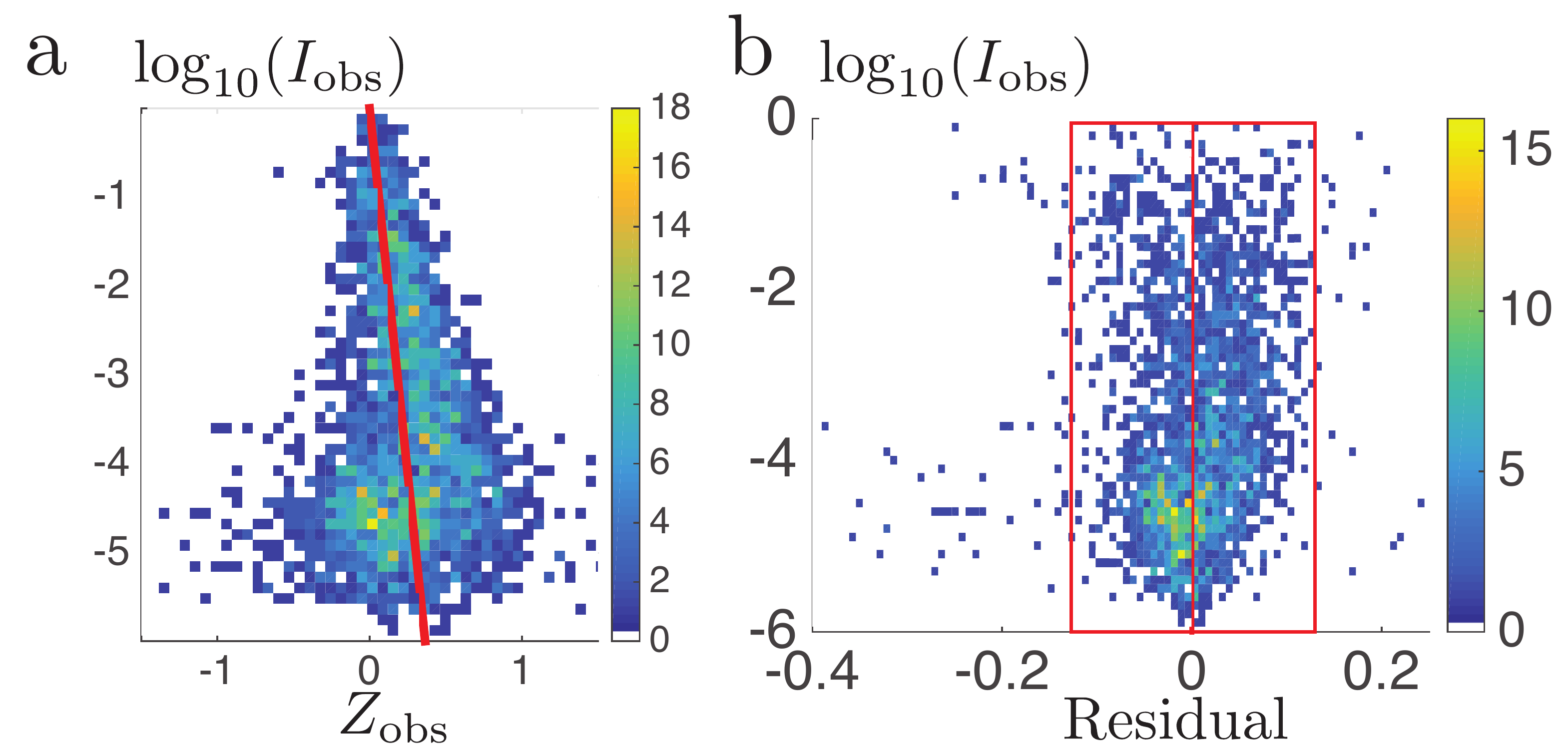}}
	\caption{(Color online) Analysis of the experiments from Ref. \cite{Martinez-marrades2014} using the Gaussian model defined in \refn{def:model}. (a) Density representation of the experimental data: measured scattered intensity $I_i$ as a function of the measure height $Z_{i}$: (colormap) experimental number of observations per boxes (red line) linear regression for $\ln(I)$ in terms of $z$. (b) Residual analysis and validation the linear Gaussian model: we verify that residual $R_i = \sqrt{I_i} Z_i + \sqrt{I_i} \ln (I_i/\widehat{I_0}_n))$ is centered (red line) and maintains a constant variance (red box) with the intensity $I_{\mathrm{obs}}$.}
	\label{fig7}
\end{figure}

\begin{figure}[t!]
	{\includegraphics[width=8.25cm]{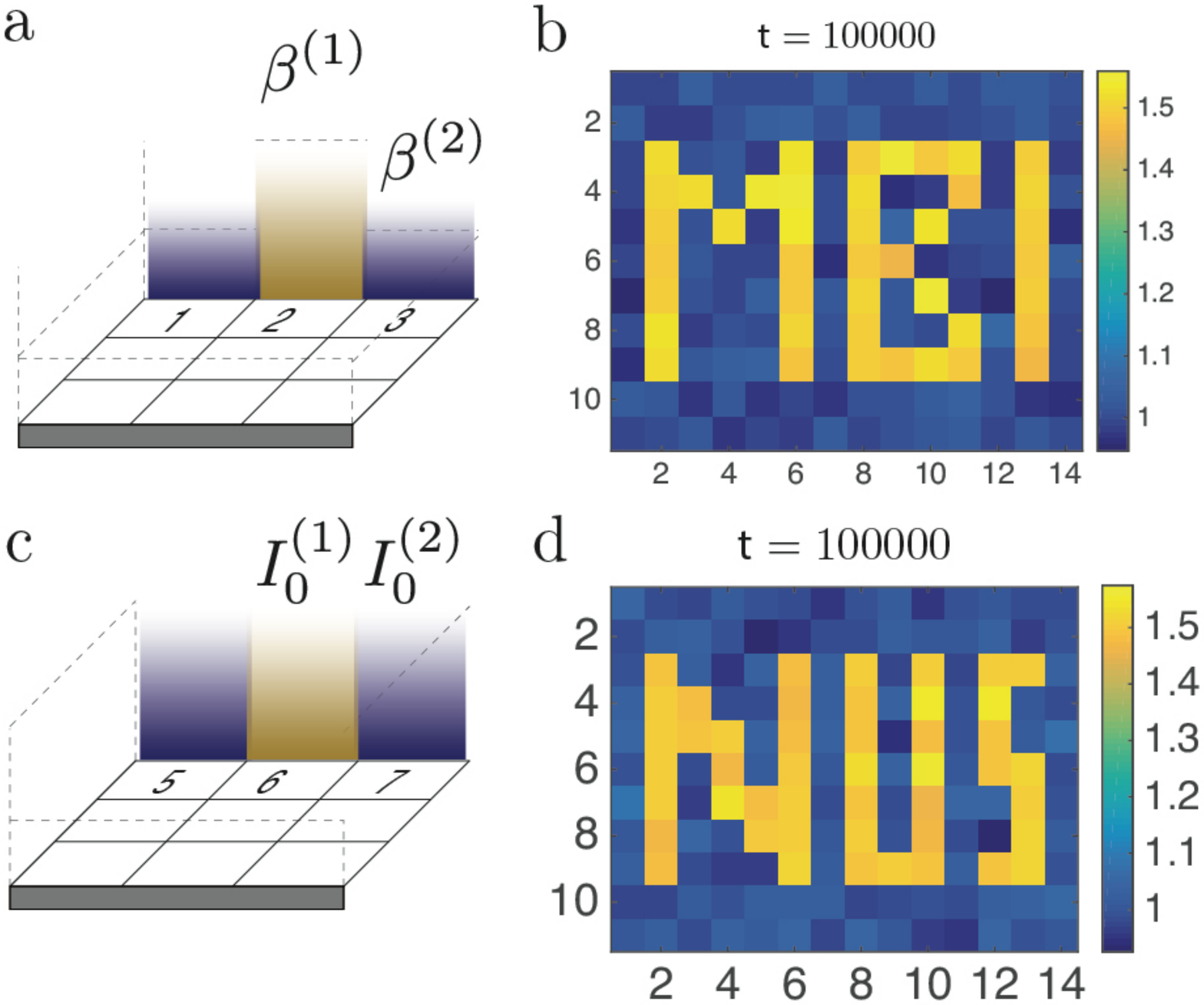}	}
	\caption{(Color online) Convergence of estimators and the imaging process of a patterned surface ($14 \times 14$ pixels) in terms of the penetration length field $\beta$ and maximal intensity $I_0$. (a) The penetration length is $\beta = \beta^{(2)}$ by default and $\beta = \beta^{(1)} = 1.50 \beta^{(2)}$ within specific pixels (MBI pattern). (b) Convergence of the estimator $\widehat{\beta}_n$, defined in \refn{eq:estimatorbetan}. (c) The maximal intensity $I_0 = I_0^{(1)}$ or $I_0 = I_0^{(2)} = 0.75 I_0^{(1)}$  within specific pixels (NUS pattern). (d) Convergence of the estimator $\widehat{I_0}_n$ defined in \refn{eq:estimatorI0}. }
	\label{Fig8}
\end{figure}

In this section, we justify that a large number of observations is required per pixel to obtain a reliable measurements of the electromagnetic field considered in Ref. \cite{Martinez-marrades2014}. We recall that the experimental data consists in $2792$ measurements of heights and intensities $(\Zi, \Ii)$ ($i \leq 2792$). We model the noise on the height measurement through the following linear Gaussian model:
\begin{align} \label{def:model}
{\Ii} \Zi = -\beta \sqrt{\Ii} \ln\left(\Ii/I_0\right)  + \sigma \eta_i,
\end{align} 
where $\eta$ is a standard Gaussian white noise process. We define the vector of unknown parameters $B = \beta (-1, \ln I_0)$, where $\beta$ refers to the penetration length and $I_0$ to the intensity at the surface, as well as $Y = \sqrt{I_i} Z_i $ (vector of observations) and $X = (\sqrt{I_i}  \ln I_i, \sqrt{I_i})$ (explicative matrix). In terms of these quantities, the model defined in \refn{def:model} reads $Y = X B + \sigma \eta$, which corresponds to the well-known linear Gaussian model. The estimator of the vector $B$ that maximizes the likelihood function is $ \widehat{B}_n =  (X^{T} X)^{-1} X^{T} Y$ \cite{Rivoirard2012} ; the developed expression corresponds to the two estimators
\begin{align}  \label{eq:estimatorbetan}
	\widehat{\beta}_n  &= - C \left\lbrace \ov{I} \cdot \ov{Z I \ln I} - \ov{I \ln I} \cdot \ov{Z I} \right\rbrace, \\
\widehat{I_0}_n  &= \exp\left( \frac{1}{\hat{\beta}_n } \left\lbrace \ov{I} \cdot \ov{Z I \ln I} - \ov{I \ln I} \cdot \ov{Z I} \right\rbrace\right),  \label{eq:estimatorI0}
\end{align}
where $\overline{I} = \sum^{n}_{i=1} I_i$ and  $C =\left(\ov{I} \cdot \ov{I \ln^2 I}- \ov{I \ln I}^2\right)^{-1}$.  The variance of the noise is also an unknown variable that can be evaluated by the estimator: 
\begin{align}
\widehat{\sigma}^2_n  = (\sum^{n}_{i = 1} R^2_i)/(n-2),
\end{align}
where $R_i = \sqrt{I_i} Z_i + \sqrt{I_i} \ln (I_i/\widehat{I_0}_n))$ is called the residual. For a Gaussian distribution of noise, we expect to have: $	\widehat{\sigma}^2_n/\sigma^2 \sim \mathcal{\nbpixel}\left(1, 1/\sqrt{n-2}\right)
$ in the limit $n \gg 1$.

We now define confidence intervals for the estimators defined in Eqs. (\ref{eq:estimatorbetan}--\ref{eq:estimatorI0}). We consider a risk level $\alpha = 0.05$: with a $1-\alpha = 0.95$ probability, the quantity $\beta$ lies within the confidence interval $\mathcal{C}(\widehat{\beta}_n)$:
\begin{align} \label{eq:confidenceinterval}
\beta \in \mathcal{C}(\widehat{\beta}_n) = \left[\widehat{\beta}_n \pm \widehat{\sigma}_n t^{(n-2)}_{1-\alpha/2}  \sqrt{\ov{I} C} \right],
\end{align}
where $t^{(n-2)}_{0.975}$ is the one-sided quantile of the Student distribution, with $t^{(n-2)}_{0.975} = 1.96 \ldots $ in the limit $n \gg 1$. As the quantity $C$ is inversely proportional to the number of observations $n$, the confidence interval \refn{eq:confidenceinterval} narrows on $\beta$ with a $1/\sqrt{n}$ speed as $n$ increases. 

From \refn{eq:confidenceinterval}, we obtain an estimate of $r$, i.e. the minimal number $n = r$ of observations required so that the estimator of $\beta$ has an error lesser than $10$\%, with probability $95\%$ probability. From experimental values, we find that the confidence interval for $\beta$ is $60,1 \pm 7.3 \, \mathrm{nm}$ for $n = 2792$. 

Therefore, the estimate for the minimal observation per pixel $r$ should around $4000$.

We now consider a system of $\nbpixel$ identical pixels. The probability that all the estimators $\widehat{\beta}^{(j)}_n$, where $1 \leq j \leq \nbpixel$ is the pixel label, are precise at $10\%$ to the exact value $\beta$ is
$\mathbb{P}\left[ \forall j,  \beta \in CI(\widehat{\beta}^{(j)}_n) \right] = (\mathbb{P}\left[ \beta \in CI(\widehat{\beta}^{(1)}_n) \right])^\nbpixel$.
We set $\mathbb{P}\left[ \forall j,  \beta \in CI(\widehat{\beta}^{(j)}_n) \right] = 1 - \epsilon$ where $\epsilon$ is the accepted risk level (in the following $\epsilon = 0.05$). Therefore, compared to the case of single pixel, the risk level on a single pixel is to be divided by a factor $\nbpixel$:  $\mathbb{P}\left[ \beta \in CI(\widehat{\beta}^{(1)}_n) \right] = 1 - 0.05/\nbpixel$. The confidence interval from \refn{eq:confidenceinterval} holds provided that $\alpha = 0.05/\nbpixel$, which corresponds to a narrower confidence interval compared to the single pixel case. However the quantity $t^{(n-2)}_{1-\epsilon/(2\nbpixel)}$  increases weakly with the number of pixels $\nbpixel$ (e.g. $t^{(\infty)}_{1-0.05/(2\times 10)} \approx 2.80$ and $t^{(\infty)}_{1-0.05/(2\times 100)} \approx 3.40$). 

In conclusion, increasing the number of pixels $\nbpixel$ does not significantly increase the required number of observations per pixel $r$.


We now apply our results to the detection of a pattern in the electromagnetic field (see Fig.  SI. \ref{Fig8}), in a simulated experiment. We consider a local electromagnetic field that takes the form of \refn{eq:io}, and in which $\beta$ and $I_0$ may take either of two values. In Fig.  SI. \ref{Fig8}, the error of the estimators $\beta$ is lower than $10 \%$ ; the image indeed results from the superposition of $t = 10^{6}$ frames, a number in agreement with the predicted threshold from \refn{eq:gaussian_approximate}: $\nbpixel \times r \approx 10^{6}$, where $\nbpixel = 14 \times 14$ and $r = 4000$.

\end{document}